\documentclass[a4paper,11pt]{article}
\usepackage{jinstpub} 
\usepackage{lineno}
\usepackage{diagbox}
\usepackage{color}
\usepackage{soul}
\usepackage{makecell}

\title{\boldmath Characterization of Multiple Channels Room Temperature Readout Electronics for Large Transition-Edge Sensor Array}

\author[a]{N. Li,}
\author[a]{X. Ren}
\author[b]{H. Gao,}
\author[b]{Z. Zhang,}
\author[b]{Y. Zhang,}
\author[b]{C. Liu,}
\author[a]{H. Li,}
\author[b,1]{and Z. Li
\note{Corresponding author.}}

\affiliation[a]{
Key Laboratory of Particle Physics and Particle Irradiation (MOE), Institute of Frontier and Interdisciplinary Science, Shandong University, Qingdao, Shandong, 266237, China
}
\affiliation[b]{
Key Laboratory of Particle Astrophysics, Institute of High
Energy Physics, CAS, 19B Yuquan Road, Shijingshan District,
100049, Beijing, China
}

\emailAdd{lizw@ihep.ac.cn}

\abstract
{
Transition-edge sensor (TES) is a highly sensitive device that is capable of detecting extremely low levels of energy.
It is characterised by low noise performance and high energy resolution.
A mature method for reading out TES signal is through time-division multiplexing (TDM) direct current superconducting quantum interference device (SQUID).
In a TDM system, the signal readout chain represents a significant source of noise other than the TES intrinsic noise. 
The noise generated by TES is in the range of several tens to several hundreds of $pA/\sqrt{Hz}$. 
In order to ensure the high energy resolution of TES, it is necessary to ensure that the noise contribution from the room temperature readout electronics is less than $10$ $pA/\sqrt{Hz}$ above 100 $Hz$. 
In this work, we have designed a low-noise, high-resolution  room temperature readout circuit for TDM. 
The equivalent current noise contribution of ADC is about $0.05$ $pA/\sqrt{Hz}$ above 100 $Hz$ and $0.46$ $pA\sqrt{Hz}$ under 30:1 multiplexing.
The resolution of the analog to digital converter (ADC) is larger than 11.5 bits, which can reconstruct the TES signal without distortion.
The readout board, which has eight channels, has JESD204B serial ports, which has greatly simplified the space of room temperature electronics. 
The readout chain is based on multi-threaded CPU processing and can transfer data at 2 $Gbps$ for each channel in real time. 
This readout board can be used in a TDM system with smaller size for large TES arrays.
}


\begin{document}
\maketitle
\flushbottom

\section{Introduction}
\label{sec:intro}
TES micro-calorimeters and bolometers have ultra low thermal noise and ultra high energy resolution.
Therefore TESs are widely used in particle physics experiments, such as X-ray~\cite{smith2024development}, $\gamma$-ray~\cite{6416013}, dark matter~\cite{BILLARD2024116465}, neutrinos~\cite{Rajteri2020}, cosmic microwave background ~\cite{Ghosh_2022} detection and so on.
For example, in X-ray detection experiments, the energy resolution is on the order of a few electronvolts ($eV$) ~\cite{smith2024development,bandler2023line}. 
In axion or dark photon detection, the energy resolution is on the order of sub-$eV$~\cite{BILLARD2024116465,TES_meV_1}.
One way to readout the TES signal is based on a direct current SQUID~\cite{Wuwentao_2022}. 
The signal is fully amplified by SQUID at cryogenic refrigerator, which provides enhanced immunity to interference and thus allows for matching with the room temperature readout circuit.
\begin{figure}[htbp]
    \centering
    \includegraphics[width=.9\textwidth]{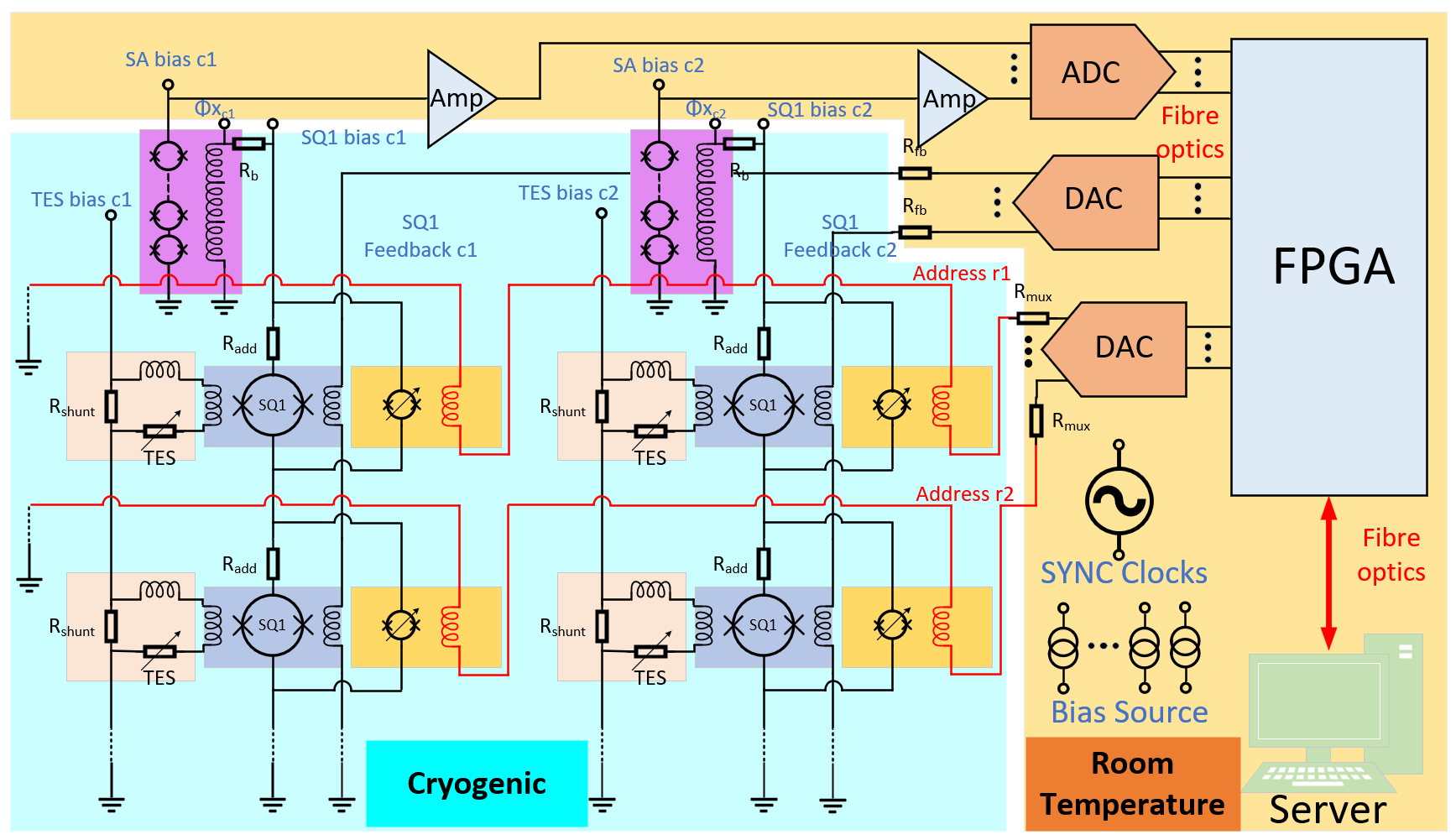}
    \caption{TDM architecture (a $2\times2$ array example here). The TES array is read out by the first-stage SQUID (SQ1) to the SQUID array (SA), and then amplified by low-noise amplifiers at room temperature. 
    The warm readout electronics consist of low-noise pre-amplifiers, high-speed AD/DA converters, signal processing FPGA, server and all bias sources for adjusting the working points of TES or SQUIDs. 
    Each row of TES array is selected alternately by $Address$ signals with time $t_{row}$.   
    }
    \label{fig:TES_readout}
\end{figure}
One of the most mature DC-SQUID readout techniques is time-division multiplexing (TDM) electronics, as illustrated in Figure~\ref{fig:TES_readout}, based on digital flux feedback (DFB) locked loops~\cite{reintsema2003prototype,Wu_2022}.
Current TDM electronics are capable of reading out more than 30 or even 100 TES units, at sample rate greater than 50 MSPS, on a single channel~\cite{battistelli2008functional,doriese2016developments, prele2016128}. 
The majority of DFB boards utilised in contemporary TDM systems employ ADCs for communication with the field-programmable gate array (FPGA) in parallel digital signal readout.
This approach requires a significant amount of space on a printed circuit board (PCB).
In order to control the size of PCB, one DFB board usually carries only one ADC chip with one readout channel. 
The number of detectors determines the number of DFB boards required for large-scale TES array experiments. 
For example, 720 TES detectors require 24 DFB boards according to the 30:1 multiplexing ratio. 
This greatly increases the size and complexity of the electronics design.
This work presents a room temperature readout circuit based on an 8-channel ADC with a 125 MSPS sampling rate.
At a 30:1 multiplexing ratio, only 3 DFB boards are required, each comprising 240 TES detectors.
Furthermore, the ADC digital link is based on the JESD204B high-speed serial protocol, which does not significantly expand the design size of a DFB board.
This can notably reduce the size of the TDM electronics system.

In order to avoid any impact on the energy resolution of the TES, it is essential to ensure that the room temperature electronics noise contribution does not represent the primary source of system noise.
Therefore, it is desirable that the noise contribution of the room temperature readout circuit does not exceed $10$ $pA/\sqrt{Hz}$.
Furthermore, the singal-to-noise ratio (SNR) of TES is generally less than $60$ $dB$~\cite{kilbourne2007uniform, 5643112}, which can be calculated based on the amplitude of signal and the noise. The equivalent the effective bit resolution (ENOB) of $60$ $dB$ is 9.67 bits, which is based on $ENOB=(SNR-1.76)/6.02$. 
Thus the ENOB of ADC should be larger than 9.67 bits.
This paper presents comprehensive tests to validate the performance of this room temperature readout circuit.

The paper is organized as follows: 
Section~\ref{sec:sys} presents digital readout circuit design, the two stage SQUID characterization and testing system configuration;
Section~\ref{sec:sys_pf} presents the SQUID $V-\phi$ curve, the noise contribution and resolution of ADC;
and Section~\ref{sec:conl} concludes the paper.

\section{System design}
\label{sec:sys}
\subsection{Digital cards design}
\label{sec:dcd}
\begin{figure}[htbp]
    \centering
    \includegraphics[width=0.9\textwidth]{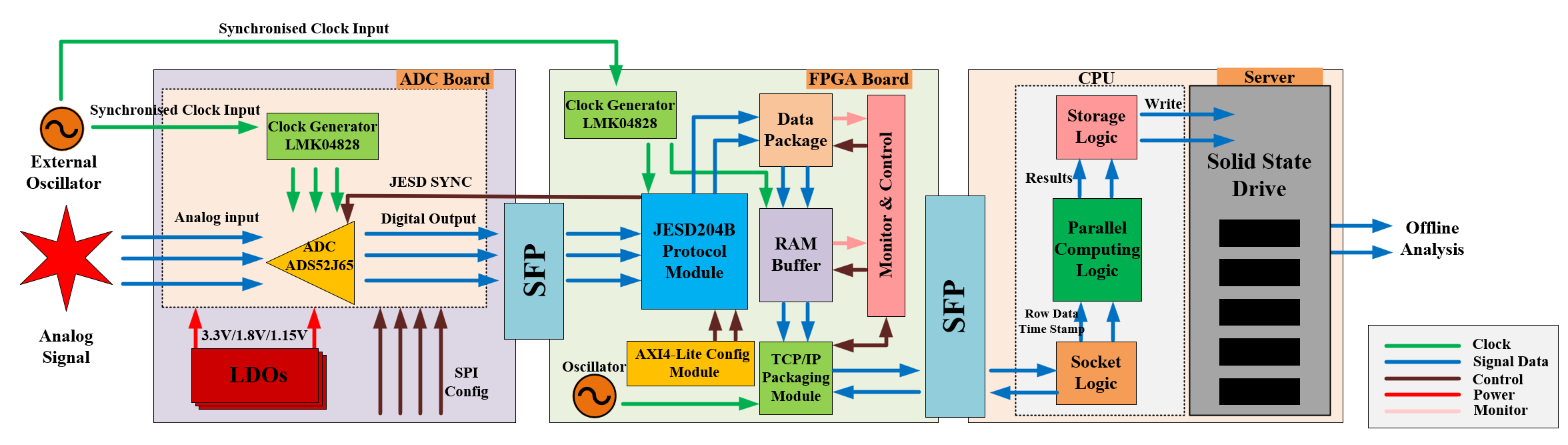}
    \caption{Digital readout circuit schematic design.
    The analogue signal represents the output signal from the preamplifier.
    An external homogeneous crystal provides a homogeneous clock for the ADC board and the FPGA board. 
    The RAM buffer in FPGA and the host computer's multi-threaded logic ensure that the signals are transferred in real time without data loss.}
    \label{fig:sys_principle}
\end{figure}

The principal component of the digital cards include the ADC board and the digital signal processing FPGA board. 
The essential configuration of these elements and digital signal communication links, is illustrated in Figure ~\ref{fig:sys_principle}.
The ADC board incorporates a TI's analogue-to-digital converter (ADC) ADS52J65, which is capable of supporting up to eight channels of SQUID signal readout.
The ADC operates at a sampling rate of 125 $MHz$, which can support 60 samples to calculate the feedback signal when $t_{row}=500$ $ns$. 
TES need have an energy resolution of at least $2eV@5.9keV$ with $0.4$ $eV$ energy bin width, which necessitates the counts of ADC code at least 14750.
The digital code number of this ADC is up to 16 bits, which is large enough for $0.4$ $eV$ energy bin width.
The JESD204B transmission protocol is supported.
In order to guarantee the accuracy of signal transmission and synchronisation in real-time data processing systems, the ADC board design synchronous clock generation module based on TI's ultra-low noise clock module LMK04828.
The SYNC/SYSREF signals, as specified in the JESD204B protocol, are generated by an internal cascaded phase locked loop (PLL) and integrated voltage controlled oscillator (VCO), resulting in a synchronous clock output.
All input reference clocks within the system are supplied by the same external 10 $MHz$ signal generator.
The device power supply is provided by a low-noise linear voltage regulator (LDO) module.

The digital signal processing board has been designed with the Xilinx XCKU060 FPGA serving as the central processing unit.
The ADC board is connected to the FPGA board via four $10$ $Gbps$ SFP optical ports, which facilitate electrical isolation and thereby enhance the noise performance of the ADC board.
At the ADC's default sampling rate (125 MSPS), the theoretical data rate between the two boards is up to $20$ $Gbps$, including the eight ADC analogue channels.
The data transmission between the FPGA and the host computer is based on the $40$ $Gbps$ transmission control protocol/internet protocol (TCP/IP)~\cite{easynet}.
To prevent the packet loss during the signal upload process, the host computer receives the data in parallel via two CPU threads and stores it in the shared memory.
Subsequently, the data is sorted into packets by the other two CPU threads in real time.
Ultimately, the fifth thread facilitates the online storage of these packets on the hard disk.
Each packet comprises 4096 bytes, with a 64-byte header.
The use of RAM as a cache in FPGA can also help avoiding packet loss due to delay in data uploading process.
Furthermore, a digital control board was designed based on Xilinx XC7A100T. 
This board supports 300 IO control ports compatible with different level standards and can provide SPI configuration signals or other control signals for ADC and signal processing boards, as illustrated in Figure~\ref{fig:sys}. 

\subsection{Testing system}
The entire TES readout system should be comprised of two components: the cryogenic electronics and the room temperature readout electronics as illustrated in Section~\ref{sec:intro}.
The multi-stage series-connected structure of SQUID exhibits exceptionally large flux conversion coefficients, which effectively mitigate the noise contribution of the room-temperature readout electronics~\cite{Wuwentao_2022,Drung_2006}.
These essential SQUID design parameters are tested in our laboratory, as illustrated in Table~\ref{tab:SQUID}.
\begin{table}[htbp]
    \centering
    \caption{Two-stage DC-SQUID transfer parameters\label{tab:SQUID}}
    \begin{tabular}{c|c}
        \hline
        $Parameter$ & $Value$ \\
        \hline
        Mutual inductance of input coil of Input SQUID $1/M_{in,IS}$ & {$8$ ${\mu}A/\phi_0$} \\
        Conversion coefficient $\left|dI_{IS,0}/d\phi_{in,IS,0}\right|$ of Input SQUID at lock point & $45$ {${\mu}A/\phi_0$} \\
        Mutual inductance of input coil of SQUID Array $1/M_{in,SA}$ & {$28$ ${\mu}A/\phi_0$} \\
        Conversion coefficient $\left|dV_{SA,0}/d\phi_{in,SA,0}\right|$ of SQUID Array at lock point & $10$ {$mV/\phi_0$} \\
        \hline
    \end{tabular}
\end{table} 

\begin{figure}[htbp]
    \centering
    \includegraphics[height=0.36\textwidth]{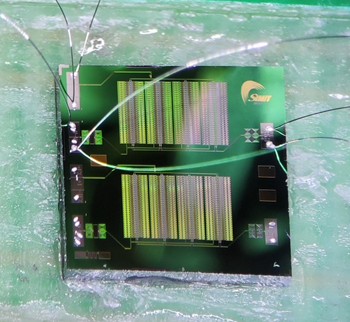}
    \quad
    \includegraphics[height=0.36\textwidth]{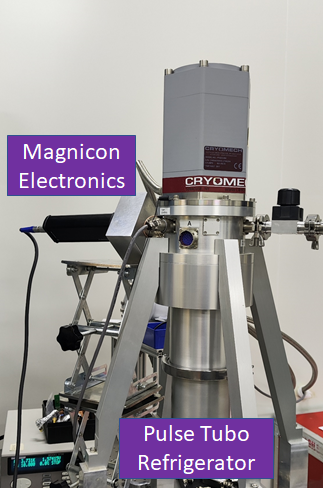}
    \quad
    \includegraphics[height=0.3\textwidth]{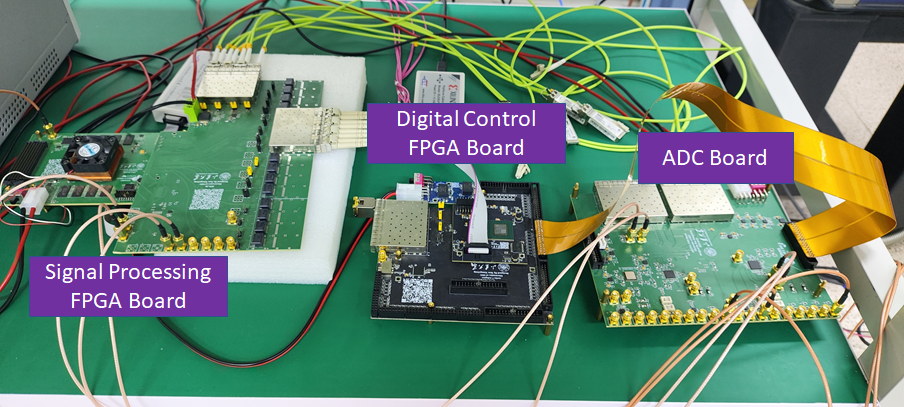}
    \caption{SQUID array test system. The up-left figure is the bonding status of the SQUID array and placement.
    The up-right figure is the interconnection status between cryogenic system and Magnicon electronics. The bottom figure shows the digital cards.}
    \label{fig:sys}
\end{figure}

Although there are two-stage SQUIDs in TDM system, only one SQUID array is enough for verifying the functionality of the readout circuit and analyze the noise contribution of ADC.
The SQUID array was tested in a pulse tubo refrigerator at 4 $K$ temperature, as illustrated in Figure~\ref{fig:sys}. 
The SQUID array comprises 100 SQUID units in series, which can provide further signal amplification.
To prevent ground loops, the SQUID ground was connected to the room temperature electronics ground at a single point.

The room temperature readout system comprises commercially available Magnicon electronics and digital readout boards, as detailed in Section~\ref{sec:dcd}. 
The Magnicon electronics provide bias source for SQUID array, enabling flux working (lock) point adjustment. 
The ADC input signal and the Magnicon output signal are coupled through a 50 $\Omega$ coaxial cable to prevent the interference of electromagnetic fields and ensure impedance matching.

\section{System performance}
\label{sec:sys_pf}

\subsection{SQUID flux response curve and noise analysis}
\label{sec:SA}
\begin{figure}[htbp]
    \centering
    \includegraphics[height=0.5\textwidth]{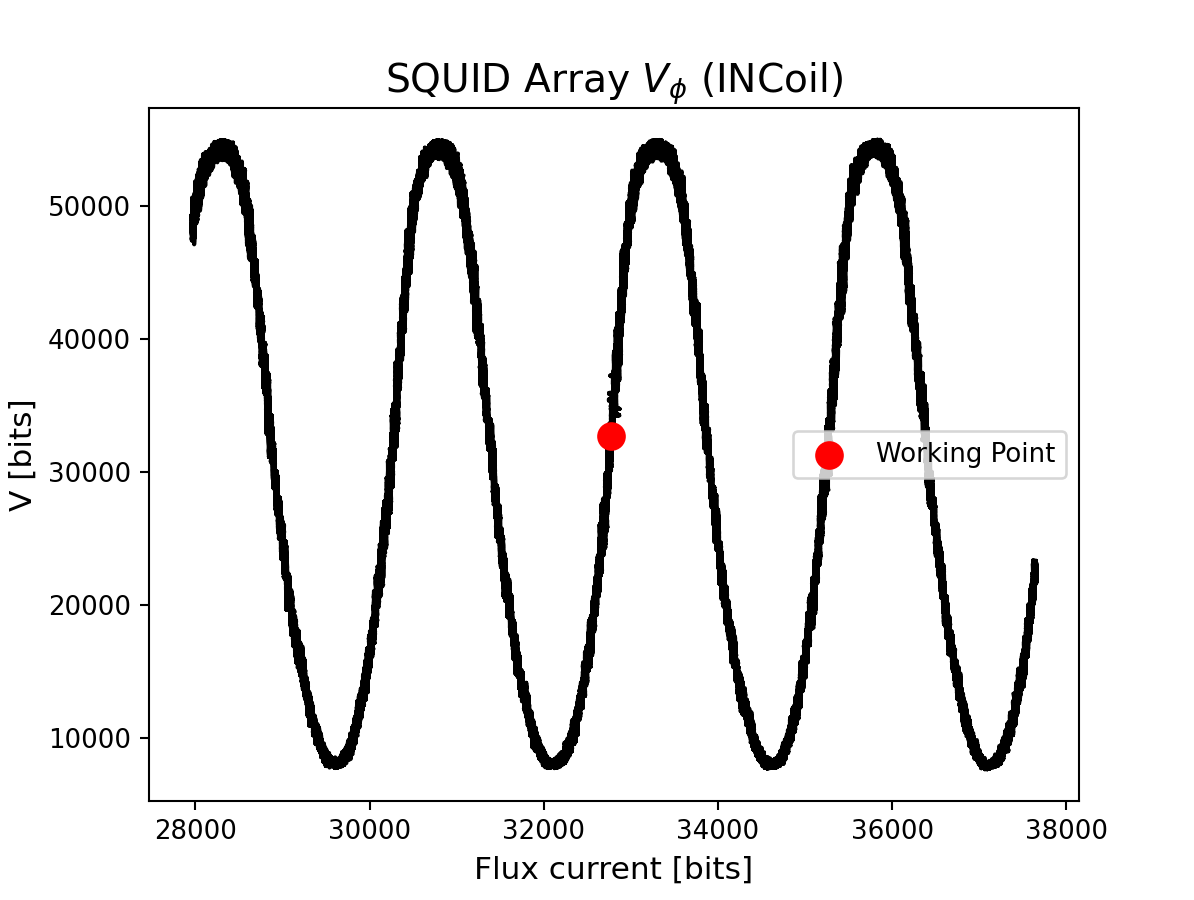}

    \caption{Flux response curve of SQUID array input coil at zero working point. The red point is the zero working point. The x and y coordinates indicate the ADC sampling digits. The flux current of input coil and voltage of SQUID is captured by channels 3 and 8, respectively.}
    \label{fig:SQUID_Vphi}
\end{figure}

To verify the functionality of the readout circuit, we capture the $V$-$\phi$ curve of SQUID array, as illustrated in Figure~\ref{fig:SQUID_Vphi}.
System noise is a key parameter in the determination of the energy resolution of TES detectors.
In order to analyze the room-temperature electronic noise contribution, the $V$-$\phi$ curve of SQUID must be adjusted to pass the zero point, the red working point in Figure~\ref{fig:SQUID_Vphi}, ensuring that a zero-flux input is equivalent to a zero-voltage output.
This process is achieved through the utilization of the AMP mode of Magnicon electronics and readout based on ADC board.
All data is captured at $4$ $Gbps$ sampling rate without data loss.

The noise of the cryogenic and room-temperature electronics readout channels, in addition to the TES itself, constitutes the primary component of the system noise.
As outlined in section~\ref{sec:intro}, the equivalent flux noise of the TES detector tested in a laboratory setting is of the order of approximately tens of $pA/\sqrt{Hz}$.
It is therefore essential to ensure that the noise of the SQUID and room-temperature readout electronics on the signal readout link does not exceed $10$ $pA/\sqrt{Hz}$.
The signal transfer function of the system should be satisfied as following:
\begin{equation}
\label{eq:1}
    \begin{aligned}
        V_{out} = I_{in,IS}M_{in,IS}\left|\frac{dI_{IS,0}}{d\phi_{in,IS,0}}\right|M_{in,SA}\left|\frac{dV_{SA,0}}{d\phi_{in,SA,0}}\right|G
    \end{aligned}
\end{equation}
$V_{out}$ denotes the output voltage following room temperature electronics amplification, which may be considered the input voltage signal of the ADC. 
The current signal of the Input SQUID input coil is represented by $I_{in,IS}$. 
The value of $G=1100$ denotes the gain of the Magnicon electronics.  

\begin{figure}[htbp]
    \centering
    \includegraphics[height=0.6\textwidth]{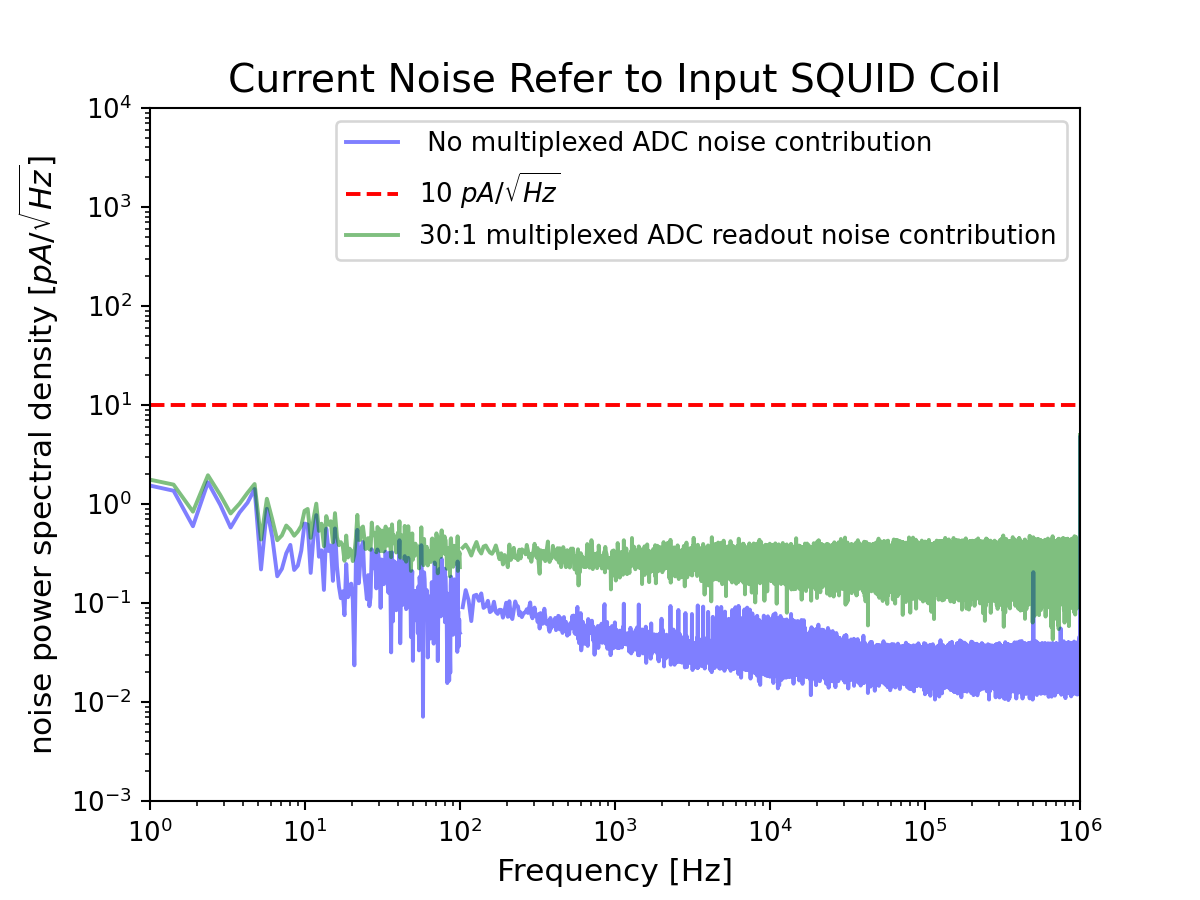}
    \caption{ADC noise contribution analysis. 
    The red line represents the 10 $pA/\sqrt{Hz}$ noise density refer to input coil of input SQUID. 
    The blue line depicts the ADC current noise contribution $S_{I,ADC}$ of the 8th channel, which was measured by inserting a $50$ $\Omega$ terminal at the ADC input.
    The green line shows the 30:1 multiplexed noise contribution $S_{I,MUX}$ of ADC.}
    \label{fig:Noise}
\end{figure}

The system bandwidth becomes smaller after TDM multiplexing, which in turn gives rise to the aliasing of Nyquist white noise~\cite{doriese2016developments}.
In order to avoid the crosstalk caused by TDM, the row selection time $t_{row}$ for each line of TES array to be read out can be set as $t_{row} = 2{\pi}{\tau}_{OL}$~\cite{DORIESE2006808}.
${\tau}_{OL}$ is the time constant relative to the open-loop bandwidth ${f}_{OL}=1/(2{\pi}{\tau}_{OL})$ when we model the TDM as a single-pole, low-pass filter.
Thus the integrated effective noise bandwidth of TDM is ${f}_{eff}=1/(4{\tau}_{OL})$. 
The Nyquist bandwidth of TDM is ${f}_{Nyq}=1/(2N_{rows}t_{row})$. 
After considering ${f}_{eff}$ and ${f}_{Nyq}$,
we can get the relationship between the number of multiplexed rows $N_{rows}$ and the aliased white noise should be satisfied:
\begin{equation}
\label{eq:2}
    S_{I,MUX,W}  
    \begin{cases}
        = S_{I,ADC,W}\sqrt{\frac{f_{eff}}{f_{Nyq}}} = S_{I,ADC,W}\sqrt{{\pi}N_{rows}}&, N_{rows}>1 \\
        = S_{I,ADC,W}&, N_{rows}=1
    \end{cases}
\end{equation}
$S_{I,MUX,W}$ denotes the total aliased current white noise contribution of the ADC after multiplexing.
$S_{I,ADC,W}$ denotes the single-pixel ADC current white noise contribution.
The total system noise $S_{I,MUX} = S_{I,ADC} + S_{I,MUX,W}-S_{I,ADC,W}$ when $N_{rows}>1$ is illustrated in Figure~\ref{fig:Noise}.
$S_{I,ADC}$ denotes the single-pixel ADC total current noise contribution, which inclues the non-linear $1/f$ noise and white noise $S_{I,ADC,W}$.
This figure shows that the readout circuit does not account for the major noise component. 
The noise contribution of  8th ADC is about $0.05$ $pA\sqrt{Hz}$ and the 30:1 multiplexed noise is about $0.46$ $pA\sqrt{Hz}$ above $100$ $Hz$, which is less than $10$ $pA\sqrt{Hz}$. 
The noise contribution on four randomly selected ADC channels are presented in Table~\ref{tab:nois}.
\begin{table}[htbp]
    \centering
    \caption{ADC noise contribution results\label{tab:nois}}
    \resizebox{0.95\textwidth}{0.12\textwidth}{
    \begin{tabular}{c|c|c|c|c}
        \hline
         Channel number & $1$ & $2$ & $5$ & $8$\\
        \hline
        Intrinsic voltage noise ($nV\sqrt{Hz}$) & \thead{$194.31@100Hz$ \\ $85.59@100kHz$} & \thead{$152.65@100Hz$ \\ $70.79@100kHz$} & \thead{$112.56@100Hz$ \\ $47.50@100kHz$} & \thead{$116.29@100Hz$ \\ $53.95@100kHz$} \\
        \hline
        Current noise contribution ($pA\sqrt{Hz}$) & \thead{$0.09@100Hz$ \\ $0.04@100kHz$} & \thead{$0.07@100Hz$ \\ $0.03@100kHz$} & \thead{$0.05@100Hz$ \\ $0.02@100kHz$} & \thead{$0.05@100Hz$ \\ $0.02@100kHz$} \\
        \hline
        30:1 TDM current noise contribution ($pA\sqrt{Hz}$) & \thead{$0.28@100Hz$ \\ $0.23@100kHz$} & \thead{$0.26@100Hz$ \\ $0.22@100kHz$} & \thead{$0.24@100Hz$ \\ $0.21@100kHz$} & \thead{$0.24@100Hz$ \\ $0.21@100kHz$}\\
        \hline
    \end{tabular}
    }
\end{table}
The $2\times$ gain of the DC-coupled amplifier at the front end of channel 5 and 8 of the ADC.
All results are less than 1 $pA/\sqrt{Hz}$ above 100 $Hz$.

\subsection{Resolution}

\begin{figure}[htbp]
    \centering
    \includegraphics[height=0.6\textwidth]{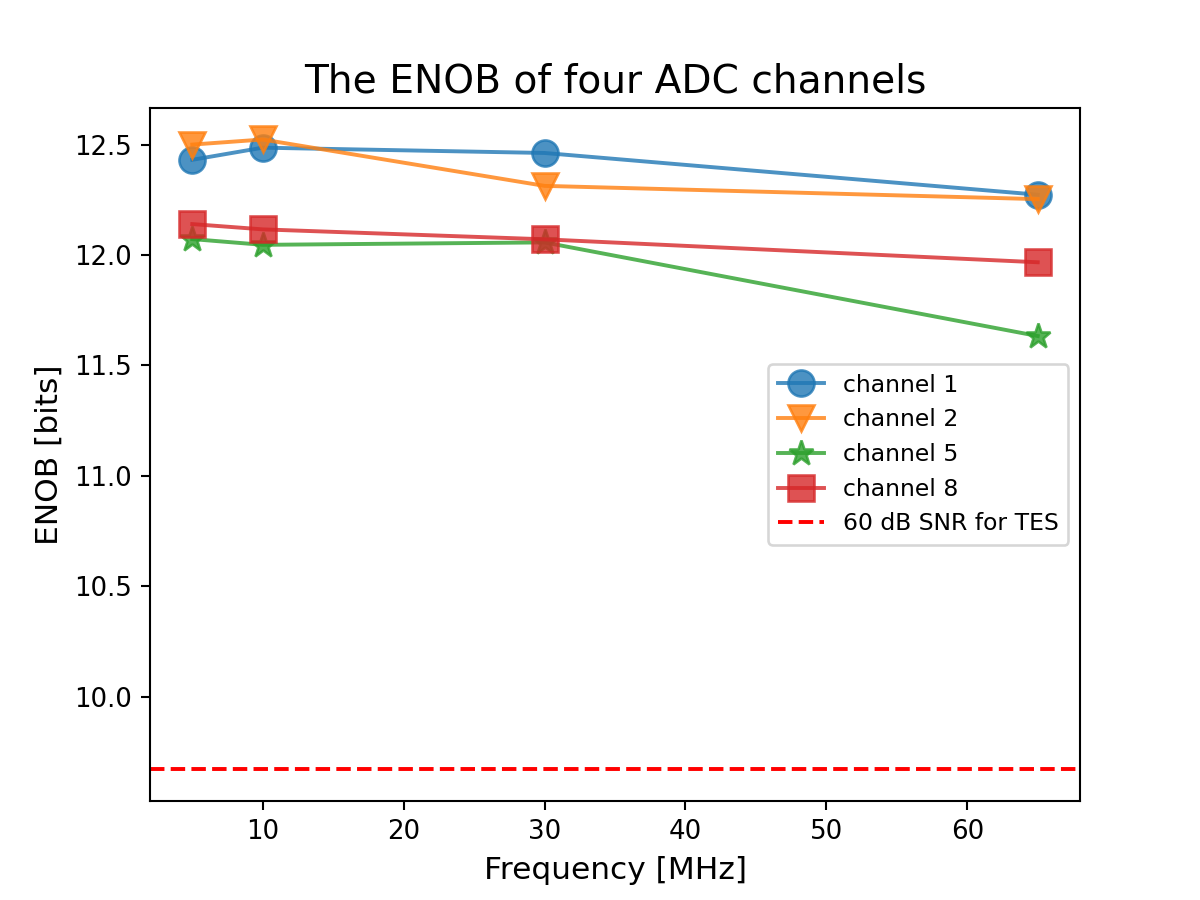}
    \caption{The ENOB of four ADC channels. 
    The four solid lines with different colors and markers represent the measured ENOB of four channels. 
    The red dased line represents the equivalent energy resolution of TES.}
    \label{fig:ENOB}
\end{figure}

ADCs inherently introduce quantization noise, making it impossible to achieve the ideal 16-bit digital resolution in practice. A key parameter for assessing signal reconstruction quality is the Effective Number of Bits (ENOB) of the ADC. To achieve a signal-to-noise ratio (SNR) of 60 dB, the TES system requires an equivalent resolution of 9.67 bits. The ENOB of four different ADC channels was evaluated using sinusoidal wave signals of varying frequencies, as shown in Figure~\ref{fig:ENOB}.
All measured results exceed 11.5 bits, which comfortably meet the required performance criteria.

\section{Conclusion}
\label{sec:conl}
An 8-channel low-noise, high-resolution room temperature readout circuit has been designed. 
Real-time data uploading, sorting and storing at a single channel data rate of 2 $Gbps$ is achieved by using the multi-threaded CPU of host computer.
The noise density contribution of digital readout electronics can be reduced to a level of approximately $0.46$ $pA/\sqrt{Hz}$ under 30:1 multiplexing.
The resolution of ADC is greater than 9.67 bits. This ensures the successful reconstruction of the TES signal.
It is possible to readout more TES units with smaller PCB size.
We will design a small-space TDM system with greater than 30:1 multiplexing ratios based on this readout circuit in future.


\acknowledgments
This work is supported by the National Key Research
and Development Program of China 
(Grant 
No.2021YFC2203400,
No.2021YFC2203402,
No.2022YFC2204900, 
No.2022YFC2204904, 
No.2020YFC2201704), the National Natural Science Foundation of China (Grant No.12105157), Youth Innovation Promotion Association CAS 2021011 and Scientific Instrument Developing Project of the Chinese Academy of Sciences,Grant No.YJKYYQ20190065.


\bibliographystyle{JHEP}
\bibliography{biblio}

\providecommand{\href}[2]{#2}\begingroup\raggedright\begin{thebibliography}{10}

\bibitem{smith2024development}
S.J.~Smith, J.S.~Adams, S.R.~Bandler, R.B.~Borrelli, J.A.~Chervenak, F.A.~Colazo-Petit et~al., \emph{Development of the microcalorimeter detector for the athena/x-ray integral field unit},  in \emph{Space Telescopes and Instrumentation 2024: Ultraviolet to Gamma Ray}, vol.~13093, pp.~247--262, SPIE, 2024.

\bibitem{6416013}
S.~Hatakeyama, M.~Ohno, R.M.T.~Damayanthi, H.~Takahashi, Y.~Kuno, K.~Maehata et~al., \emph{Development of hard x-ray and gamma-ray spectrometer using superconducting transition edge sensor}, \href{https://doi.org/10.1109/TASC.2013.2241386}{\emph{IEEE Transactions on Applied Superconductivity} {\bfseries 23} (2013) 2100804}.

\bibitem{BILLARD2024116465}
J.~Billard, J.~Gascon, S.~Marnieros and S.~Scorza, \emph{Transition edge sensors with sub-ev resolution and cryogenic targets (tesseract) at the underground laboratory of modane (lsm)}, \href{https://doi.org/https://doi.org/10.1016/j.nuclphysb.2024.116465}{\emph{Nuclear Physics B} {\bfseries 1003} (2024) 116465}.

\bibitem{Rajteri2020}
M.~Rajteri, M.~Biasotti, M.~Faverzani, E.~Ferri, R.~Filippo, F.~Gatti et~al., \emph{Tes microcalorimeters for ptolemy}, \href{https://doi.org/10.1007/s10909-019-02271-x}{\emph{Journal of Low Temperature Physics} {\bfseries 199} (2020) 138}.

\bibitem{Ghosh_2022}
S.~Ghosh, Y.~Liu, L.~Zhang, S.~Li, J.~Zhang, J.~Wang et~al., \emph{Performance forecasts for the primordial gravitational wave detection pipelines for alicpt-1}, \href{https://doi.org/10.1088/1475-7516/2022/10/063}{\emph{Journal of Cosmology and Astroparticle Physics} {\bfseries 2022} (2022) 063}.

\bibitem{bandler2023line}
S.R.~Bandler, J.S.~Adams, E.G.~Amatucci, E.R.~Canavan, J.A.~Chervenak, R.S.~Cumbee et~al., \emph{Line emission mapper microcalorimeter spectrometer}, {\emph{Journal of Astronomical Telescopes, Instruments, and Systems} {\bfseries 9} (2023) 041002}.

\bibitem{TES_meV_1}
R.K.~Romani, Y.-Y.~Chang, R.~Mahapatra, M.~Platt, M.~Reed, I.~Rydstrom et~al., \emph{A transition edge sensor operated in coincidence with a high sensitivity athermal phonon sensor for photon coupled rare event searches}, \href{https://doi.org/10.1063/5.0234265}{\emph{Applied Physics Letters} {\bfseries 125} (2024) 232601}.

\bibitem{Wuwentao_2022}
W.~Wu, Z.~Lin, Z.~Ni, P.~Li, T.~Liang, G.~Zhang et~al., \emph{Development of series squid array with on-chip filter for tes detector}, \href{https://doi.org/10.1088/1674-1056/ac2b91}{\emph{Chinese Physics B} {\bfseries 31} (2022) 028504}.

\bibitem{reintsema2003prototype}
C.D.~Reintsema, J.~Beyer, S.W.~Nam, S.~Deiker, G.C.~Hilton, K.~Irwin et~al., \emph{Prototype system for superconducting quantum interference device multiplexing of large-format transition-edge sensor arrays}, {\emph{Review of Scientific Instruments} {\bfseries 74} (2003) 4500}.

\bibitem{Wu_2022}
X.~Wu, Q.~Yu, Y.~He, J.~Liu and W.~Chen, \emph{Multiplexing technology based on squid for readout of superconducting transition-edge sensor arrays}, \href{https://doi.org/10.1088/1674-1056/ac693c}{\emph{Chinese Physics B} {\bfseries 31} (2022) 108501}.

\bibitem{battistelli2008functional}
E.S.~Battistelli, M.~Amiri, B.~Burger, M.~Halpern, S.~Knotek, M.~Ellis et~al., \emph{Functional description of read-out electronics for time-domain multiplexed bolometers for millimeter and sub-millimeter astronomy}, {\emph{Journal of Low Temperature Physics} {\bfseries 151} (2008) 908}.

\bibitem{doriese2016developments}
W.B.~Doriese, K.M.~Morgan, D.A.~Bennett, E.V.~Denison, C.P.~Fitzgerald, J.W.~Fowler et~al., \emph{Developments in time-division multiplexing of x-ray transition-edge sensors}, {\emph{Journal of low temperature physics} {\bfseries 184} (2016) 389}.

\bibitem{prele2016128}
D.~Pr{\^e}le, F.~Voisin, M.~Piat, T.~Decourcelle, C.~Perbost, C.~Chapron et~al., \emph{A 128 multiplexing factor time-domain squid multiplexer}, {\emph{Journal of Low Temperature Physics} {\bfseries 184} (2016) 363}.

\bibitem{kilbourne2007uniform}
C.A.~Kilbourne, S.R.~Bandler, A.-D.~Brown, J.A.~Chervenak, E.~Figueroa-Feliciano, F.M.~Finkbeiner et~al., \emph{Uniform high spectral resolution demonstrated in arrays of tes x-ray microcalorimeters},  in \emph{UV, X-Ray, and Gamma-Ray Space Instrumentation for Astronomy XV}, vol.~6686, pp.~52--61, SPIE, 2007.

\bibitem{5643112}
F.M.~Finkbeiner, C.N.~Bailey, S.R.~Bandler, R.P.~Brekosky, A.D.~Brown, J.A.~Chervenak et~al., \emph{Development of embedded heatsinking layers for compact arrays of x-ray tes microcalorimeters}, \href{https://doi.org/10.1109/TASC.2010.2091237}{\emph{IEEE Transactions on Applied Superconductivity} {\bfseries 21} (2011) 223}.

\bibitem{easynet}
Z.~He, D.~Korolija and G.~Alonso, \emph{Easynet: 100 gbps network for hls},  in \emph{2021 31st International Conference on Field-Programmable Logic and Applications (FPL)}, (Los Alamitos, CA, USA), pp.~197--203, IEEE Computer Society, sep, 2021, \href{https://doi.org/10.1109/FPL53798.2021.00040}{DOI}.

\bibitem{Drung_2006}
D.~Drung, C.~Hinnrichs and H.-J.~Barthelmess, \emph{Low-noise ultra-high-speed dc squid readout electronics}, \href{https://doi.org/10.1088/0953-2048/19/5/S15}{\emph{Superconductor Science and Technology} {\bfseries 19} (2006) S235}.

\bibitem{DORIESE2006808}
W.~Doriese, J.~Beall, W.~Duncan, L.~Ferreira, G.~Hilton, K.~Irwin et~al., \emph{Progress toward kilopixel arrays: 3.8ev microcalorimeter resolution in 8-channel squid multiplexer}, \href{https://doi.org/https://doi.org/10.1016/j.nima.2005.12.146}{\emph{Nuclear Instruments and Methods in Physics Research Section A: Accelerators, Spectrometers, Detectors and Associated Equipment} {\bfseries 559} (2006) 808}.

\end{thebibliography}\endgroup







\end{document}